\documentstyle[12pt]{article}
\newcommand{\beq}{\begin{equation}}
\newcommand{\eeq}{\end{equation}}
\newcommand{\beqa}{\begin{eqnarray}}
\newcommand{\eeqa}{\end{eqnarray}}
\newcommand{\ba}{\begin{array}}
\newcommand{\ea}{\end{array}}

\begin{document}

\begin{center}
{\large \bf Chaos and Quantum Chaos in Nuclear Systems}
\footnote{Presented to the VI Workshop
"Perspectives on Theoretical Nuclear Physics", 12--14 October 1995,
Cortona (Italy).}
\end{center}

\vskip 1. truecm

\begin{center}
{\bf Luca Salasnich}
\footnote{Work in collaboration with J.M.G. Gomez and V.R. Manfredi.}
\vskip 0.5 truecm
Dipartimento di Fisica "G. Galilei" dell'Universit\`a di Padova, \\
Via Marzolo 8, I 35131 Padova, Italy \\
and\\
Departamento de Fisica Atomica, Molecular y Nuclear \\
Facultad de Ciencias Fisicas, Universidad "Complutense" de Madrid, \\
Ciudad Universitaria, E 28040 Madrid, Spain\\
\end{center}

\vskip 1. truecm

\begin{center}
{\bf Abstract}
\end{center}

\vskip 0.5 truecm
\par
The presence of chaos and quantum chaos is shown in
two different nuclear systems. We analyze the chaotic
behaviour of the classical SU(2) Yang--Mills--Higgs system,
and then we study quantum chaos in the nuclear shell model calculating
the spectral statistics of $A=46$--$50$ atomic nuclei.

\vskip 1. truecm

\par
{\bf 1. Introduction}
\vskip 0.5 truecm
\par
A classical system is {\it chaotic} if has exponential
divergence of initially closed trajectories in the phase space,
i.e. small differences in the initial conditions
produce great changes in the final behaviour.
The phenomenon of chaos, which has become very popular, rejuvenated
interest in nonlinear dynamics.
Through numerical simulations on modern computers the presence of chaos
has been discovered to be pervasive in many dynamical systems of physical
interest [1].
\par
In quantum mechanics one cannot apply classical concepts and methods
directly being the notion of trajectory absent. Nevertheless, many
efforts have been made to establish the features of quantum systems
which reflect the qualitative difference in the behaviour of their
classical counterparts [2].
\par
The {\it quantum chaos}, or better the {\it quantum chaology} [3],
studies the properties of quantum systems which are classically chaotic.
The energy fluctuation properties of systems with underlying
classical chaotic behaviour and time--reversal symmetry agree with
the predictions of the Gaussian Orthogonal Ensemble (GOE) of
random matrix theory, whereas quantum analogs of classically
integrable systems display the characteristics
of the Poisson statistics [2--4].
\par
In this paper we analyze the presence of chaos and quantum chaos
in two systems of interest for nuclear physics. In the first part
we study analytically the suppression of classical chaos in the spatially
homogenous SU(2) Yang--Mills--Higgs (YMH) system.
In the second part we analyze numerically quantum chaos in the
nuclear shell model by studying the spectral statistics
of low--lying states of $A=46$--$50$ atomic nuclei.

\vskip 0.5 truecm
\par
{\bf 2. Chaos in the SU(2) Yang--Mills--Higgs system}
\vskip 0.5 truecm
\par
In the last years there has been much interest in the chaotic behaviour
of classical field theories [5]. Usually the order--chaos transition
in these systems has been studied
numerically by using Lyapunov exponents and Poincar\`e sections [1,2].
Less attention has been paid to analytical criteria [6,7].
Here we study the spatially homogenous SU(2) YMH system.
Obviously, the constant field approximation implies that our SU(2) YMH
system is a toy model for classical non--linear dynamics,
with the attractive feature that the model emerges from particle physics.
\par
The lagrangian density for the SU(2) YMH system is given by [8]:
\beq
L=-{1\over 4}F_{\mu \nu}^{a}F^{\mu \nu a}
+{1\over 2}(D_{\mu}\phi )^+(D^{\mu}\phi )
-V(\phi ) \; ,
\eeq
where:
\beq
F_{\mu \nu}^{a}=\partial_{\mu}A_{\nu}^{a}-\partial_{\nu}A_{\mu}^{a}+
g\epsilon^{abc}A_{\mu}^{b}A_{\nu}^{c} \; ,
\eeq
\beq
(D_{\mu}\phi )=\partial_{\mu}\phi - i g A_{\mu}^b T^b\phi
\; ,
\eeq
with $T^b=\sigma^b/2$, $b=1,2,3$, generators of the SU(2) algebra,
and where the potential of the scalar field (the Higgs field) is:
\beq
V(\phi )=\mu^2 |\phi|^2 + \lambda |\phi|^4 \; .
\eeq
We work in the (2+1)--dimensional Minkowski space ($\mu =0,1,2$) and
choose spatially homogenous Yang--Mills and the Higgs fields:
\beq
\partial_i A^a_{\mu} = \partial_i \phi = 0, \;\; i=1,2
\eeq
i.e. we consider the system in the region in which space fluctuations of
fields are negligible compared to their time fluctuations.
\par
In the gauge $A^a_0=0$ and using the real triplet representation for the
Higgs field we obtain:
$$
L={1\over 2}({\dot {\vec A_1}}^2+{\dot {\vec A_2}}^2)+
{\dot {\vec \phi}}^2
-g^2 [{1\over 2}{\vec A_1}^2 {\vec A_2}^2
-{1\over 2} ({\vec A_1} \cdot {\vec A_2})^2+
$$
\beq
+({\vec A_1}^2+{\vec A_2}^2){\vec \phi}^2
-({\vec A_1} \cdot {\vec \phi})^2 -({\vec A_2} \cdot {\vec \phi})^2]
-V({\vec \phi}),
\eeq
where ${\vec \phi}=(\phi^1,\phi^2,\phi^3)$,
${\vec A_1}=(A_1^1,A_1^2,A_1^3)$, and
${\vec A_2}=(A_2^1,A_2^2,A_2^3)$.
\par
When $\mu^2 >0$ the potential $V$ has a minimum in $|{\vec \phi}|=0$,
but for $\mu^2 <0$ the minimum is:
$$
|{\vec \phi_0}|=({-\mu^2\over 4\lambda })^{1\over 2}=v
$$
which is the non zero Higgs vacuum. This vacuum is degenerate
and after spontaneous symmetry breaking the physical vacuum can be
chosen ${\vec \phi_0} =(0,0,v)$.
If $A_1^1=q_1$, $A_2^2=q_2$ and the other components of the
Yang--Mills fields are zero.
The system hamiltonian in the Higgs vacuum is:
\beq
H={1\over 2}(p_1^2+p_2^2)
+g^2v^2(q_1^2+q_2^2)+{1\over 2}g^2 q_1^2 q_2^2 \; ,
\eeq
where $p_1={\dot q_1}$ and $p_2={\dot q_2}$. Obviously $w^2=2 g^2v^2$ is the
mass term of the Yang--Mills fields.
\par
The transition order--chaos in systems with two
degrees of freedom may be studied by the curvature criterion of
potential energy [6]. It is however important to point out that
{\it in general} the curvature
criterion guarantees only a {\it local instability} and should therefore
be combined with the Poincar\`e sections [9].
\par
At low energy the motion near the minimum of the potential:
\beq
V(q_1,q_2)=g^2 v^2 (q_1^2+q_2^2)+{1\over 2} g^2 q_1^2 q_2^2,
\eeq
where the curvature is positive, is periodic or quasiperiodic and is
separated from the instability region by a line of zero curvature;
if the energy is increased, the system will be for some initial conditions
in a region of negative curvature, where the motion is chaotic.
According to this scenario, the energy of order$\to$chaos transition
$E_c$ is equal to the minimum value of the line of zero gaussian
curvature $K(q_1 ,q_2 )$ on the potential--energy surface.
For our potential the gaussian curvature vanishes at the points
that satisfy the equation:
\beq
{\partial^2 V \over \partial q_1^2}
{\partial^2 V \over \partial q_2^2}-
({\partial^2 V \over \partial q_1 \partial q_2})^2=
(2g^2v^2 +g^2 q_2^2)(2g^2v^2+g^2q_1^2)-4g^4 q_1^2q_2^2=0.
\eeq
It is easy to show that the minimal energy on the
zero--curvature line is given by:
\beq
E_c=V_{min}(K=0,\bar{q_1})=6 g^2 v^4,
\eeq
and occurs at $\bar{q_1}=\pm \sqrt{2}v$.
Therefore the chaos--order transition depends on the parameter
$\epsilon = g^2 v^4/E$: for $0 < \epsilon < 6$ a relevant region of the
phase--space is chaotic, while for $\epsilon > 6$ the system
becomes regular. This result shows that
the Higgs field value in the vacuum $v$ plays an important role:
for large values it makes the system regular
in agreement with previous numerical calculations of Savvidy [5].
Also the Yang--Mills coupling constant $g$ has the
same role. Instead for fixed $v$ and $g$ there
is an order--chaos transition increasing the energy $E$ [10].

\vskip 0.5 truecm
\par
{\bf 3. Quantum chaos in the nuclear shell model}
\vskip 0.5 truecm
\par
One of the best systems for the study of quantum chaos is the atomic
nucleus, which has been the subject of many investigations [11].
In atomic nuclei, the fluctuation properties of experimental energy levels are
best studied in the domain of neutron and proton resonances near the nucleon
emission threshold, where a large number of levels with the same spin and
parity in the same nucleus are present, and an excellent agreement with GOE
predictions has been found [12].
\par
In this work we undertake the statistical analysis of the shell--model
energy levels in the $A=46$--$50$ region.
By using second--quantization notation,
the nuclear shell--model hamiltonian may be written as [13]:
\beq
H=\sum_{\alpha} \epsilon_{\alpha}a_{\alpha}^+a_{\alpha}
+\sum_{\alpha \beta \gamma \delta} <\alpha \beta|V|\delta \gamma >
a_{\alpha}^+ a_{\beta}^+ a_{\gamma} a_{\delta},
\eeq
where the labels denote the accessible single--particle states,
$\epsilon_{\alpha}$ is the corresponding single--particle energy,
and $<\alpha \beta|V|\delta \gamma >$ is the two--body matrix element
of the nuclear residual interaction.
\par
Exact calculations are performed with the $f_{7/2}$, $p_{3/2}$, $f_{5/2}$,
and $p_{1/2}$ single--particle states, assuming a $^{40}$Ca inert core.
The diagonalizations are performed in the {\it m}--scheme using a fast
implementation of the Lanczos algorithm with the code ANTOINE [14].
For a fixed number of valence protons and neutrons
we calculate the energy spectrum for
projected total angular momentum $J$ and total isospin $T$.
The interaction we use is a minimally modified Kuo--Brown
realistic force with monopole improvements [15].
\par
We calculate the $T=T_z$ states from $J=0$ to $J=9$
for all the combinations of $6$ active nucleons,
i.e. $^{46}$V, $^{46}$Ti, $^{46}$Sc, and $^{46}$Ca, and also
for $^{48}$Ca and $^{50}$Ca.
\par
Since we are looking for deviations from chaotic features, we are
mainly interested in the low--lying levels up to a few MeV above the $JT$
yrast line. For each $JT$ set of levels the spectrum is
mapped into unfolded levels with quasi--uniform level density
by using the constant temperature formula [16].
\par
The spectral statistic $P(s)$ is used
to study the local fluctuations of the energy levels. $P(s)$ is
the distribution of nearest--neighbour spacings
$s_i=({\tilde E}_{i+1}-{\tilde E}_i)$ of the unfolded levels ${\tilde
E}_i$. It is obtained by accumulating the number of spacings that lie within
the bin $(s,s+\Delta s)$ and then normalizing $P(s)$ to unity.
\par
For quantum systems whose classical analogs are integrable,
$P(s)$ is expected to follow the Poisson limit, i.e.
$P(s)=\exp{(-s)}$. On the other hand,
quantal analogs of chaotic systems exhibit the spectral properties of
GOE with $P(s)= (\pi / 2) s \exp{(-{\pi \over 4}s^2)}$ [2--4].
\par
The distribution $P(s)$ is the best spectral statistic to analyze
shorter series of energy levels and
the intermediate regions between order and chaos.
The $P(s)$ distribution can be compared to the Brody distribution [17]
\beq
P(s,\omega)=\alpha (\omega +1) s^{\omega} \exp{(-\alpha s^{\omega+1})},
\eeq
with
\beq
\alpha = \big( \Gamma [{\omega +2\over \omega+1}] \big)^{\omega +1}.
\eeq
This distribution interpolates between the Poisson distribution ($\omega =0$)
of integrable systems and the GOE distribution ($\omega =1$) of
chaotic ones, and thus the parameter $\omega$ can be used as a simple
quantitative measure of the degree of chaoticity.
\par
In order to obtain more meaningful statistics, $P(s)$ is calculated using the
unfolded level spacings of the whole set of $J=0$--$9$ levels
for fixed $T$ up to a given energy limit above the yrast line.
Thus the number of spacings included is reasonably large.

\vskip 0.5 truecm

\begin{center}
\begin{tabular}{|ccccccc|} \hline\hline
Energy & $^{46}$V & $^{46}$Ti & $^{46}$Sc & $^{46}$Ca & $^{48}$Ca & $^{50}$Ca
\\ \hline
$\leq 4$ MeV & 1.14 & 0.90 & 0.81 & 0.41 & 0.58 & 0.67 \\
$\leq 5$ MeV & 1.10 & 0.81 & 0.96 & 0.53 & 0.58 & 0.69 \\
$\leq 6$ MeV & 0.93 & 0.94 & 0.99 & 0.51 & 0.66 & 0.62 \\
\hline\hline
\end{tabular}
\end{center}
\vskip 0.5 truecm
{\bf Table 1}: Brody parameter $\omega$ for the nearest neighbour level
spacings distribution for $0\leq J\leq 9$, $T=T_z$ states up to
$4$, $5$ and $6$ MeV above the yrast line in the analyzed nuclei.

\vskip 0.5 truecm

\par
Table 1 shows the best fit Brody parameter $\omega$ of the $P(s)$
distribution for the $J=0$--$9$ set of
level spacings in the $A=46$ nuclei up to $4$, $5$ and $6$ MeV above
the yrast line. Clearly, $^{46}$V, $^{46}$Ti and $^{46}$Sc are chaotic
for these low energy levels, but there is a considerable deviation from GOE
predictions in $^{46}$Ca, which is a single closed--shell nucleus.
In view of the peculiarity of this nucleus, we performed
calculations for $^{48}$Ca and $^{50}$Ca, and obtained again strong deviations
toward regularity, as the values of Table 1 show.
\par
To explain these results we observe that the two--body matrix elements
of the proton--neutron interaction are, on average, larger than those of
the proton--proton and neutron--neutron interactions.
Consequently the single--particle mean--field motion in nuclei with both
protons and neutrons in the valence orbits suffers
more disturbance and is thus more chaotic.

\vskip 0.5 truecm
\par
{\bf 4. Conclusions}
\vskip 0.5 truecm
\par
The presence of chaos in the homogenous SU(2) Yang--Mills--Higgs system has
been studied. The results show that for large values of the Higgs field
in the vacuum there is a suppression of the chaotic behaviour of the
Yang--Mills fields. In the future it will be of great interest to analyze
this effect also in a non homogenous situation.
\par
In the study of the nuclear shell model we have seen that
the disturbance of single--particle motion
by the two--body interaction is greatest in light nuclei, where the size
of the range of the single--particle orbits is not much longer
than the range of the nuclear force.
In particular, for Ca isotopes, we find significant deviations
from the predictions of the
random--matrix theory which suggest that some spherical nuclei
are not as chaotic in nature as the conventional view assumes.

\vskip 0.5 truecm

\begin{center}
*****
\end{center}
\par
The author has been supported by a Fellowship form the University of Padova
and is grateful to the "Ing. Aldo Gini" Foundation of Padova
for a partial support. He acknowledges Prof. J. M. G. Gomez
for his kind hospitality at the Department of Atomic,
Molecular and Nuclear Physics of "Complutense" University.

\vskip 0.5 truecm

\parindent=0. pt
\section*{References}

\vskip 0.5 truecm

[1] A. J. Lichtenberg, M. A. Lieberman: {\it Regular and Stochastic
Motion} (Springer--Verlag, 1983)

[2] M. C. Gutzwiller: {\it Chaos in Classical and Quantum Mechanics}
(Springer--Verlag, Berlin, 1990)

[3] A. M. Ozorio de Almeida: {\it Hamiltonian Systems: Chaos and
Quantization} (Cambridge University Press, Cambridge, 1990)

[4] K. Nakamura: {\it Quantum Chaos} (Cambridge Nonlinear Science Series,
Cambridge, 1993)

[5] G. K. Savvidy: Nucl. Phys. B {\bf 246}, 302 (1984);
T. Kawabe, S. Ohta: Phys. Lett. B {\bf 334}, 127 (1994);
T. Kawabe: Phys. Lett. B {\bf 343}, 254 (1995)

[6] M. Toda: Phys. Lett. A {\bf 48}, 335 (1974)

[7] S. Graffi, V. R. Manfredi, L. Salasnich: Mod. Phys. Lett. B {\bf 7},
747 (1995)

[8] C. Itzykson, J. B. Zuber: {\it Quantum Field Theory} (McGraw--Hill, 1985)

[9] G. Benettin, R. Brambilla, L. Galgani: Physica A {\bf 87}, 381 (1977)

[10] L. Salasnich: "Chaos Suppression in the SU(2) Yang--Mills--Higgs System",
Preprint DFPD/95/TH/15, to be published in Phys. Rev. D.

[11] M. T. Lopez--Arias, V. R. Manfredi, L. Salasnich:
Riv. Nuovo Cim. {\bf 17}, N. 5, (1994) 1.

[12] R. U. Haq, A. Pandey, O. Bohigas: Phys. Rev. Lett. {\bf 48}
(1982) 1086.

[13] R. D. Lawson: {\it Theory of the Nuclear Shell Model}
(Clarendon, Oxford 1980).

[14] E. Caurier: computer code ANTOINE, C.R.N., Strasbourg (1989);
E. Caurier, A. P. Zuker, A. Poves: in {\it Nuclear Structure of
Light Nuclei far from Stability: Experiment ad Theory}, Proceedings of
the Obernai Workshop 1989, Ed. G. Klotz (C.R.N, Strasbourg, 1989).

[15] E. Caurier, A. P. Zuker, A. Poves, G. Martinez--Pinedo:
Phys. Rev. C {\bf 50} (1994) 225.

[16] G. E. Mitchell, E. G. Bilpuch, P. M. Endt,
J. F. Jr. Shriner: Phys. Rev. Lett. {\bf 61} (1988) 1473;
J. F. Jr. Shriner, G. E. Mitchell, T. von Egidy:
Z. Phys. {\bf 338} (1991) 309.

[17] T. A. Brody: Lett. Nuovo Cimento {\bf 7} (1973) 482.

[18] J. M. G. Gomez, V. R. Manfredi, L. Salasnich:
"Spectral Statistics of Large Shell Model Calculations",
to be published in Proceedings of the V International Spring Seminar
'Perspectives on Nuclear Structure', Ravello (Italy),
22--26 Maggio 1995, Ed. A. Covello (World Scientific).

\end{document}